\documentclass[11pt]{article}
\usepackage{amssymb,cite}
\makeatletter
\@addtoreset{equation}{section}
\makeatother

\def\ben{\begin{equation}}
\def\een{\end{equation}}

\let\a=\alpha \let\b=\beta \let\g=\gamma  
    
\let\l=\lambda  \let\n=\nu

\let\C=\Chi

\def\nn{\nonumber} \def\bd{\begin{document}} \def\ed{\end{document}}
\def\ds{\documentstyle} \let\fr=\frac \let\bl=\bigl \let\br=\bigr
\let\Br=\Bigr \let\Bl=\Bigl
\let\bm=\bibitem
\let\na=\nabla
\let\pa=\partial \let\ov=\overline
\newcommand{\be}{\begin{equation}}
\newcommand{\ee}{\end{equation}}
\def\ba{\begin{array}}
\def\ea{\end{array}}
\def\ft#1#2{{\textstyle{{\scriptstyle #1}\over {\scriptstyle #2}}}}
\def\fft#1#2{{#1 \over #2}}
\def\del{\partial}
\def\vp{\varphi}
\def\sst#1{{\scriptscriptstyle #1}}
\def\oneone{\rlap 1\mkern4mu{\rm l}}
\def\td{\tilde}
\def\wtd{\widetilde}
\def\ie{\rm i.e.\ }
\def\dalemb#1#2{{\vbox{\hrule height .#2pt
        \hbox{\vrule width.#2pt height#1pt \kern#1pt
                \vrule width.#2pt}
        \hrule height.#2pt}}}
\def\square{\mathord{\dalemb{6.8}{7}\hbox{\hskip1pt}}}
\newcommand{\ho}[1]{$\, ^{#1}$}
\newcommand{\hoch}[1]{$\, ^{#1}$}
\newcommand{\bea}{\begin{eqnarray}}
\newcommand{\eea}{\end{eqnarray}}
\newcommand{\ra}{\rightarrow}
\newcommand{\lra}{\longrightarrow}
\newcommand{\Lra}{\Leftrightarrow}
\newcommand{\ap}{\alpha^\prime}
\newcommand{\bp}{\tilde \beta^\prime}
\newcommand{\tr}{{\rm tr} }
\newcommand{\Tr}{{\rm Tr} }
\def\0{{\sst{(0)}}}
\def\1{{\sst{(1)}}}
\def\2{{\sst{(2)}}}
\def\3{{\sst{(3)}}}
\def\4{{\sst{(4)}}}
\def\5{{\sst{(5)}}}
\def\6{{\sst{(6)}}}
\def\7{{\sst{(7)}}}
\def\8{{\sst{(8)}}}
\def\n{{\sst{(n)}}}
\def\cA{{{\cal A}}}
\def\cB{{{\cal B}}}
\def\cF{{{\cal F}}}
\def\cH{{{\cal H}}}
\def\tV{\widetilde V}
\def\tW{\widetilde W}
\def\tH{\widetilde H}
\def\tE{\widetilde E}
\def\tF{\widetilde F}
\def\tA{\widetilde A}
\def\im{{{\rm i}}}
\def\tY{{{\wtd Y}}}
\def\ep{{\epsilon}}
\def\vep{{\varepsilon}}
\def\R{\rlap{\rm I}\mkern3mu{\rm R}}
\def\bD{{{\bar D}}}
\def\R{\rlap{\rm I}\mkern3mu{\rm R}}
\def\bD{{{\bar D}}}
\def\R{{{\mathbb R}}}
\def\C{{{\mathbb C}}}
\def\H{{{\mathbb H}}}
\def\CP{{{\mathbb C}{\mathbb P}}}
\def\RP{{{\mathbb R}{\mathbb P}}}
\def\Z{{{\mathbb Z}}}
\def\bA{{{\mathbb A}}}
\def\bB{{{\mathbb B}}}
\def\bC{{{\mathbb C}}}
\def\bD{{{\mathbb D}}}
\def\bE{{{\mathbb E}}}
\def\bZ{{{\mathbb Z}}}
\def\Re{{{\mathfrak{Re}}}}
\def\Im{{{\mathfrak{Im}}}}
\def\cosec{{\,\hbox{cosec}\,}}
\def\Gm{{\Gamma_{\!\! -}}}
\def\Gp{{\Gamma_{\!\! +}}}
\def\stan{{standard }}
\def\nonstan{{supernumerary }}
\def\FF2{{ {}_{\sst 2}F_{\sst 1} }}
\def\FFF{{ {}_{\sst 3}F_{\sst 2} }}
\def\const{\rm constant}

\newcommand{\tamphys}{\it Center for Theoretical Physics,
Texas A\&M University, College Station, TX 77843}

\newcommand{\upenn}{\it Department of Physics and Astronomy,\\ University
of Pennsylvania, Philadelphia, PA 19104}

\newcommand{\damtp}{\it DAMTP, Centre for Mathematical Sciences,
 Cambridge University,\\  Wilberforce Road, Cambridge CB3 OWA, UK}

\newcommand{\brussels}{\it Physique Th\'eorique et Math\'ematique,
Universit\'e Libre de Bruxelles,\\ Campus Plaine C.P. 231, B-1050
Bruxelles, Belgium}

\newcommand{\auth}{ G.W. Gibbons{\hoch *}, H. L\"u{\hoch {\ddagger 1}}, 
Don N. Page{\hoch {\ddagger 2}} and C.N. Pope\hoch{\ddagger 1}}

\begin{document}
\begin{flushright}

DAMTP-2004-37\ \ \
Alberta Thy 07-04\ \ \
MIFP-04-07\\
{\bf hep-th/0404008}\\
April\  2004
\end{flushright}

\vspace{10pt}

\begin{center}

{\large {\bf The General Kerr-de Sitter Metrics in All Dimensions 
            }}

\vspace{20pt}
\auth

\vspace{20pt}

\hoch{*}\damtp

\vspace{10pt}

\hoch{\ddagger}{\it George P. and Cynthia W. Mitchell
Institute for Fundamental Physics,\\ Texas A\& M University,
College Station, TX 77843-4242, USA}

%\vspace{10pt} {\hoch{\dagger}\brussels}

\vspace{40pt}

\underline{ABSTRACT}
\end{center}

   We give the general Kerr-de Sitter metric in arbitrary spacetime
dimension $D\ge 4$, with the maximal number $[(D-1)/2]$ of independent
rotation parameters.  We obtain the metric in Kerr-Schild form, where
it is written as the sum of a de Sitter metric plus the square of a
null-geodesic vector, and in generalised Boyer-Lindquist coordinates.
The Kerr-Schild form is simpler for verifying that the Einstein
equations are satisfied, and we have explicitly checked our results
for all dimensions $D\le 11$.  We discuss the global structure of the
metrics, and obtain formulae for the surface gravities and areas of
the event horizons.  We also obtain the Euclidean-signature solutions,
and we construct complete non-singular compact Einstein spaces on
associated $S^{D-2}$ bundles over $S^2$, infinitely many for each odd
$D \ge 5$.

{\vfill\leftline{}\vfill \vskip 10pt \footnoterule 
{\footnotesize {\hoch 1}
Research supported in part by DOE grant
DE-FG03-95ER40917.}

{\footnotesize
{\hoch 2}
Permanent address: Theoretical Physics Institute, 412 Physics Lab., 
University of Alberta, Edmonton,}

{\footnotesize $\phantom *$ Alberta T6G 2J1, Canada.  Research
  supported in part by the Natural Science and Engineering }

{\footnotesize $\phantom *$ Research
  Council of Canada.}
}

\pagebreak
\setcounter{page}{1}

\tableofcontents
\addtocontents{toc}{\protect\setcounter{tocdepth}{2}}
\newpage

\section{Introduction}

   It is a remarkable fact that there are roughly $10^{20}$ rotating
black holes in the observable universe, and the spacetime near each
one of them is given to a very good approximation by a simple explicit
exact solution of the Einstein vacuum equations called, after its
discoverer, the Kerr metric \cite{Kerr}.  An almost equally remarkable
fact, which played an important role in its derivation, is that, in a
certain sense, the metric $g_{\mu \nu}$ is given {\sl exactly} by its
linearised approximation around the flat metric $\eta_{\mu \nu}$.
More precisely, the metric can be cast into the so-called Kerr-Schild
\cite{KerrSchild} form:
%%%%%
\ben
ds^2 =g_{\mu \nu} dx^\mu dx^\nu  = \eta_{\mu \nu} dx^\mu dx^\nu 
 + {2 M \over U} \, (k_\mu dx^\mu)^2\,,\label{ks1}
\een
%%%%%
where $k_\mu $ is null and geodesic with respect to both the full metric
 $g_{\mu \nu}$ and the flat metric $\eta _{\mu \nu}$.

   Explicitly, in flat coordinates $x,y,z,t$, one has
%%%%%%
\ben
k=k_\mu dx^\mu =dt  +{r(xdx+ydy) +a (xdy-ydx) \over r^2+a^2} +
{zdz \over r}
\,, 
\een
%%%%%
and 
%%%%%
\ben
U=  r + \fft{a^2\, z^2}{r^3}\,,   
\een
%%%%%
where $r$ is defined\footnote{Note that $r$ is {\sl not} the usual
radial coordinate in flat spacetime.} by
%%%%%
\ben
{x^2+y^2 \over r^2 +a^2 } +{z^2 \over r^2 }=1\,.
\een
%%%%%

   Some years later, Myers and Perry \cite{MyersPerry} were able to
show that a simple generalisation of this ansatz yields the exact
solution for the metric of a rotating black hole in arbitrary
dimensions.  Explicitly, they found that, for even spacetime
dimensions $D=2n\ge 4$, the null vector is
%%%%%
\ben
k=k_\mu dx^\mu =dt  +\sum_{i=1}^{n-1} {r(x_i dx_i+y_idy_i) +
    a_i (x_idy_i-y_idx_i) \over
 r^2+a_i^2} +{zdz \over r}\,,\label{mpk} 
\een
%%%%%
with
%%%%%
\ben
U= \fft1{r}\, 
\Bigl( 1- \sum_{i=1}^{n-1}  {a_i^2\, (x^2_i +y_i^2) \over (r^2+a_i^2)^2 } 
\Bigr)\, 
\prod_{j=1}^{n-1} (r^2+a^2_j)\,,\label{mpu}
\een
%%%%%
and
%%%%%
\ben
\sum_{i=1}^{n-1} {x_i^2+y_i^2 \over r^2 +a_i^2 } 
    +{z^2 \over r^2 }=1\,.\label{reqn}
\een
%%%%%.
Substituting (\ref{mpk}) and (\ref{mpu}) into (\ref{ks1}), one obtains
the generalisation of the Ricci-flat Kerr metric in $2n$ spacetime
dimensions, with $(n-1)$ independent rotation parameters $a_i$ in
$(n-1)$ orthogonal spatial 2-planes.

   If the number of spacetime dimensions is odd, say $D=2n+1$, there
are then $n$ pairs of spatial coordinates and no $z$ coordinate, and
so the terms involving $z$ are omitted. $U$ is then $1/r$ times the
right-hand side of equation (\ref{mpu}).
  
   Shortly after Kerr's paper, Carter discovered \cite{Carter}
a generalisation of the Kerr solution describing a rotating black hole
in four-dimensional  de Sitter or anti-de Sitter backgrounds, of 
which he said \cite{Carter2}
\begin{quote} {\it Although I don't think that there is much physical 
justification
for believing a non-zero $\Lambda$ term, it is perhaps worth 
quoting the result
as a geometrical curiosity.}
\end{quote}
  
   Times have changed, and there is currently great interest in the
cosmological term, both positive and negative, and in various
spacetime dimensions.  Hawking, Hunter and Taylor-Robinson have given
\cite{hawhuntay}, without detailed derivation, a generalisation of the
five-dimensional Myers-Perry solution to include a cosmological term,
and they raised the question of what form it should take in higher
dimensions. The purpose of this article is to answer that question.

   The solutions of Carter in four dimensions, and Hawking, Hunter and
Taylor-Robinson in five dimensions, are already quite complicated, and
so one needs some guiding principle to aid one's search. An important
clue is provided by Carter's observation that his solution may also be
cast in a generalised Kerr-Schild form, in which one replaces the flat
Minkowski background metric $\eta _{\mu \nu}$ by that of de Sitter or
anti-de Sitter space. In other words, one should now write
%%%%%
\ben
g_{\mu \nu}= \bar g_{\mu \nu}+ {2M\over U}\,  k_\mu k_\nu\,,
\label{ksans}
\een 
%%%%%    
where $k_\mu$ is null with respect to both the exact metric $g_{\mu
\nu}$ and some background metric $\bar g_{\mu \nu}$.  In our case,
$\bar g_{\mu \nu}$ will be taken to be the metric of de Sitter or
anti-de Sitter spacetime.  Now quite generally, one may show that if
the null vector field $k^\mu$ (with index raised using either
metric, since both give the same result) is tangent to a null-geodesic
congruence\footnote{A congruence is a family of curves, one and only
one passing through every point of spacetime. It is null if every
tangent vector is null. It is geodesic if every curve is a geodesic.},
then the Ricci tensor of $g_{\mu\nu}$ is related to that of $\bar
g_{\mu\nu}$ by \cite{dergur}
%%%%%
\be
R^\mu{}_\nu = \bar R^\mu{}_\nu  - h^\mu{}_\rho\, \bar R^\rho{}_\nu +
\ft12 \bar \nabla_\rho\, \bar \nabla_\nu\, h^{\mu\rho} + \ft12
\bar\nabla^\rho\, 
\bar \nabla^\mu\, 
h_{\nu\rho} - \ft12 \bar \nabla^\rho\, \bar \nabla_\rho\, h^\mu{}_\nu\,,
\label{ricric}
\ee
%%%%%
where $h_{\mu\nu} = \fft{2M}{U}\, k_\mu\, k_\nu$.  (See also \cite{gugu}
and \cite{stkrmahohe}.)   Thus if $\bar
g_{\mu\nu}$ satisfies $\bar R_{\mu\nu}= \Lambda\, \bar g_{\mu\nu}$, then
the full metric $g_{\mu\nu}$ will satisfy the Einstein equations
$R_{\mu\nu}= \Lambda\, g_{\mu\nu}$ with the same cosmological
constant, provided that $h_{\mu\nu}$ satisfies the {\sl linearised}
Einstein equations with respect to the background metric $\bar g_{\mu
\nu}$.  One key reason why the ``linearised approximation'' is exact
in this case is that with $g_{\mu\nu}$ written as $g_{\mu\nu} = \bar
g_{\mu\nu} + h_{\mu\nu}$, the fact that $k_\mu$ is null implies that
the inverse metric is given exactly by $g^{\mu\nu}=\bar g^{\mu\nu} -
h^{\mu\nu}$.  It is also important to note that (\ref{ricric}) is
written with {\it mixed} components; one index up and one down.

    A key element of Kerr's construction, which reappears in the
subsequent work of Carter, Myers and Perry, and Hawking, Hunter and
Taylor-Robinson, is that the solution is most naturally given in a
special coordinate system, related to the well-known Boyer-Lindquist
coordinates. These coordinates have the property that even in the
flat-space limit of the Kerr-Myers-Perry solutions, in which the mass
$M$ vanishes but the angular velocity parameters $a_i$ of the general
solution do not, they do not reduce to standard spherical coordinates.
This is clear from the expression (\ref{reqn}) for the radial
coordinate $r$; the level sets of $r$ are {\sl ellipsoids} rather than
{\sl spheres}. In fact they are {\it ellipsoids of revolution},
meaning that they are invariant under $n$ rotations in the $n$
2-planes spanned by the coordinates $(x_i,y_i)$ . It is convenient
therefore to introduce $n$ azimuthal coordinates $\phi_i$ and $n$
latitudinal coordinates $\mu_i$, such that
%%%%%%
\ben
z_i=x_i+\im\, y_i=\mu_i \, e^{\im \,\phi_i}\,, 
\een  
%%%%%         
with $\sum_i \mu_i^2=1$.  Note that for odd spacetime dimensions, in
the limit that the rotation parameters $a_i$ vanish, the level sets
become spheres, and the coordinates $\mu_i$ label the Clifford tori of
the odd-dimensional spheres $S^{2n+1}$.  The system of ellipsoids also
becomes spherical in the special case that all rotation parameters
$a_i$ are equal.  In fact, this special case is associated with an
enhancement of the general $U(1)^n$ symmetry to $U(n)$.  It would be
interesting to know how these ellipsoidal coordinates are related to
other forms of ellipsoidal coordinates that have been introduced in
higher-dimensional Euclidean space.

   An examination of the metrics of Carter and of Hawking, Hunter and
Taylor-Robinson also reveals that these special ellipsoidal
coordinates can be extended to de Sitter or anti-de Sitter space, and
they will play an essential role in what follows.

   A conventional static form of the standard (anti)-de Sitter metric
in the case of odd spacetime dimensions $D=2n+1$, with $\bar
R_{\mu\nu}= (D-1)\, \lambda\, \bar g_{\mu\nu}$, is
%%%%%%
\ben
d\bar s^2 = -(1-\lambda y^2) dt ^2 + {dy ^2 \over 1-\lambda y^2} 
  + y^2 \sum_{k=1}^n \bigl ( 
 d \hat \mu _k^2 + \hat \mu _k^2 \, d \phi _k ^2 \bigr )\,,\label{dsstat}
\een
%%%%%
with
%%%%%
\ben
\sum_{i=1}^n \hat \mu_i^2 =1\,.
\een
%%%%%%
We now define new ``spheroidal coordinates'' $(r,\mu_i)$ according to
%%%%%%
\ben
(1+\lambda a_i^2)\,  y^2 \, \hat \mu_i^2 =(r^2 +a_i^2)\, \mu_i^2\,,
\label{hatmu}
\een  
%%%%%
where
%%%%%
\be 
\sum_{i=1}^n \mu_i^2 =1\,.\label{musum} 
\ee
%%%%%
Note that (\ref{hatmu}) and (\ref{musum}) imply that
%%%%%
\ben
y^2 = \sum_{i=1}^n { (r^2 + a_i^2 )\mu _i ^2 \over 1+\lambda a_i^2}\,. 
\een
%%%%%%
In terms of the new coordinates $(r,\mu_i)$, the de Sitter metric
(\ref{dsstat}) becomes
%%%%%
\bea
d\bar s^2 &=& -W\, (1-\lambda\, r^2)\, dt^2 + F\, dr^2 + 
 \sum_{i=1}^n \fft{r^2 + a_i^2}{1+\lambda\, a_i^2} 
\,\, (d\mu_i^2 + \mu_i^2\, d\phi_i^2)\nn\\
&&\qquad  + 
 \fft{\lambda}{W\, (1-\lambda\, r^2)}\, 
\Big( \sum_{i=1}^n \fft{(r^2 + a_i^2)\, \mu_i\, d\mu_i}{1+\lambda\, a_i^2}
   \Big)^2\,,\label{dsspher}
\eea
%%%%%
where
%%%%%
\be
W \equiv \sum_{i=1}^n \fft{\mu_i^2}{1+\lambda\, a_i^2}\,,\qquad
F\equiv \fft{r^2}{1-\lambda\, r^2}\, \, 
  \sum_{i=1}^n \fft{\mu_i^2}{r^2+a_i^2}\,.\label{WFdef0}
\ee
%%%%%
A similar construction can be given in the case that the spacetime 
dimension is even.  We shall not present it here, since it will in
any case be subsumed by our discussion in the next section.

    The de Sitter metrics written in spheroidal coordinates, as in
(\ref{dsspher}), form the basis for our construction of the general
Kerr-de Sitter metrics in arbitrary dimensions. The constants $a_i$,
which have been introduced in (\ref{hatmu}) merely as parameters in a
coordinate transformation of the standard de Sitter metric, acquire
the interpretation of genuine rotation parameters for the general
Kerr-de Sitter metrics once the square of an appropriate null vector
is added, as in (\ref{ksans}).  In $D$ dimensions there are
$[(D-1)/2]$ independent rotation parameters $a_i$, characterising the
angular momenta in $[(D-1)/2]$ orthogonal 2-planes.

\section{General Kerr-de Sitter Metrics in Kerr-Schild 
     Form}\label{kerrschildsec}

   We begin by presenting our general results for the Kerr-de Sitter
metrics, expressed in Kerr-Schild form.  To do so, we introduce
$n=[D/2]$ coordinates $\mu_i$, which are subject to the constraint
%%%%%
\be
\sum_{i=1}^n \mu_i^2 =1\,,
\label{constraint}
\ee
%%%%%
together with $N=[(D-1)/2]$ azimuthal angular coordinates $\phi_i$,
the radial coordinate $r$, and the time coordinate $t$.  When the
total spacetime dimension $D$ is odd, $D=2n+1=2N+1$, there are $n$
azimuthal coordinates $\phi_i$, each with period $2\pi$.  If $D$ is
even, $D=2n=2N+2$, there are only $N=(n-1)$ azimuthal coordinates
$\phi_i$, which we take to be $(\phi_1,\phi_2,\ldots,\phi_{n-1})$.
When $D$ is odd, all the $\mu_i$ lie in the interval $0\le \mu_i\le
1$, whereas when $D$ is even, the $\mu_i$ all lie in this interval
except $\mu_n$, for which $-1\le \mu_n\le 1$.

   The Kerr-de Sitter metrics $ds^2$ satisfy the Einstein equation
%%%%%
\be
R_{\mu\nu} = (D-1)\, \lambda\, g_{\mu\nu}\,.
\ee
%%%%%
We first make the definitions
%%%%%
\be
W \equiv \sum_{i=1}^n \fft{\mu_i^2}{1+\lambda\, a_i^2}\,,\qquad
F\equiv \fft{r^2}{1-\lambda\, r^2}\, \, 
  \sum_{i=1}^n \fft{\mu_i^2}{r^2+a_i^2}\,.\label{WFdef}
\ee
%%%%%

   We find that in $D=2n+1$ dimensions the Kerr-de Sitter metrics are
given by
%%%%%
\be
ds^2 = d\bar s^2 + \fft{2M}{U}\, (k_\mu\, dx^\mu)^2\,,\label{kdsodd}
\ee
%%%%%
where the de Sitter metric $d\bar s^2$, the null one-form $k_\mu$, and the
function $U$ are given by
%%%%%
\bea
d\bar s^2 &=& - W\,(1-\lambda \, r^2)\, 
dt^2 + F\, dr^2 + \sum_{i=1}^n \fft{r^2 + a_i^2}{1+\lambda\, a_i^2} 
\,\, (d\mu_i^2 + \mu_i^2\, d\phi_i^2)\nn\\
&&\qquad  + 
 \fft{\lambda}{W\, (1-\lambda\, r^2)}\, 
\Big( \sum_{i=1}^n \fft{(r^2 + a_i^2)\, \mu_i\, d\mu_i}{1+\lambda\, a_i^2}
   \Big)^2\,,\label{oddcase}\\ 
k_\mu\, dx^\mu &=& W\, dt +  F\, dr - 
\sum_{i=1}^n \fft{a_i\, \mu_i^2}{1+\lambda\, a_i^2} \,  d\phi_i \,,
\label{oddcase2}\\
U &=& \sum_{i=1}^n \fft{\mu_i^2}{r^2 + a_i^2}\, 
\prod_{j=1}^n (r^2 + a_j^2)\,.\label{oddcase3}
\eea
%%%%%
Note that the null vector corresponding to the null one-form is
%%%%%
\be
k^\mu\, \del_\mu = -\fft{1}{1-\lambda\, r^2}\, \fft{\del}{\del t} 
    + \fft{\del}{\del r} - \sum_{i=1}^n\, \fft{a_i}{r^2+ a_i^2}\, 
\fft{\del}{\del \phi_i}\,.
\ee
%%%%%

   In $D=2n$ dimensions, we find that the Kerr-de Sitter metrics are
given by
%%%%%
\be
ds^2 = d\bar s^2 + \fft{2M}{U}\, (k_\mu\, dx^\mu)^2\,,\label{kdseven}
\ee
%%%%%
where the de Sitter metric $d\bar s^2$, the null vector $k_\mu$, and the
function $U$ are now given by
%%%%%
\bea
d\bar s^2 &=& - W\,(1-\lambda \, r^2)\, 
dt^2 + F\, dr^2 + \sum_{i=1}^n \fft{r^2 + a_i^2}{1+\lambda\, a_i^2} 
\,\, d\mu_i^2 + \sum_{i=1}^{n-1} \fft{r^2 + a_i^2}{1+\lambda\, a_i^2}
\, \, \mu_i^2\, d\phi_i^2 \nn \\
&& \qquad + 
 \fft{\lambda}{W\, (1-\lambda\, r^2)}\, 
\Big( \sum_{i=1}^n \fft{(r^2 + a_i^2)\, \mu_i\, d\mu_i}{1+\lambda\, a_i^2}
   \Big)^2\,,\label{evencase}\\ 
k_\mu\, dx^\mu &=& W\, dt +  F\, dr - 
\sum_{i=1}^{n-1} \fft{a_i\, \mu_i^2}{1+\lambda\, a_i^2} \,  d\phi_i \,,
\label{evencase2}\\
U &=& r\, \sum_{i=1}^n \fft{\mu_i^2}{r^2 + a_i^2}\, 
\prod_{j=1}^{n-1} (r^2 + a_j^2)\,.\label{evencase3}
\eea
%%%%%
In this even-dimensional case, where there is no azimuthal coordinate
$\phi_n$, there is also no associated rotation parameter, and so $a_n=0$.
In this case $k^\mu$ is given by
%%%%%
\be
k^\mu\, \del_\mu = -\fft{1}{1-\lambda\, r^2}\, \fft{\del}{\del t} 
    + \fft{\del}{\del r} - \sum_{i=1}^{n-1}\, \fft{a_i}{r^2+ a_i^2}\, 
\fft{\del}{\del \phi_i}\,.
\ee
%%%%%
The vector field $k^\mu$ is tangent to a null-geodesic congruence in both
even and odd dimensions.

    We have been led to the above expressions for the Kerr-Schild
forms of the Kerr-de Sitter metrics by casting the previously-known
$D=4$ and $D=5$ Kerr-de Sitter metrics in Kerr-Schild form, and making
the natural generalisations that are suggested by symmetries and
general considerations of covariance.  Checking the correctness of the
conjectured forms is a mechanical, if somewhat involved, procedure in
any given dimension.  We have explicitly verified, by means of
computer calculations of (\ref{ricric}) using {\it Mathematica}, that
the Kerr-de Sitter metrics we have presented here do indeed satisfy
the Einstein equations in all dimensions up to and including 11. It is
clear from the structure of the metrics that no special features arise
that would distinguish the $D\le 11$ cases from the general case, and
thus we can be confident that our expressions are valid in all
dimensions.

\section{Kerr-de Sitter Metrics in Boyer-Lindquist Coordinates}

   For some purposes it is useful to write the Kerr-de Sitter metrics
in Boyer-Lindquist coordinates.  Unlike the Kerr-Schild formulation
that we used in section \ref{kerrschildsec}, in the Boyer-Lindquist
coordinates there are no cross-terms between $dr$ and the other
coordinate differentials.  This simplifies the analysis of the event
horizons and the causal structure of the metrics.

    Let us first consider odd spacetime dimensions, $D=2n+1$.  In
Boyer-Lindquist coordinates the Kerr-de Sitter metrics are given by
%%%
\bea
ds^2 &=& - W\, (1 -\lambda\, r^2)\, 
d\tau^2 + \fft{U\, dr^2}{V-2M} +
\fft{2M}{U}\Bigl(d\tau - \sum_{i=1}^n \fft{a_i\, \mu_i^2\, d\varphi_i}{
1 + \lambda\, a_i^2}\Bigr)^2 \nn\\
&& 
+ \sum_{i=1}^n \fft{r^2 + a_i^2}{1 + \lambda\, a_i^2}
\, [d\mu_i^2 + \mu_i^2\, (d\varphi_i -\lambda\, a_i\, d\tau)^2] \nn\\
&&+
\fft{\lambda}{W\, (1-\lambda r^2)}
\Big( \sum_{i=1}^n \fft{(r^2 + a_i^2)\mu_i\, d\mu_i}{
1 + \lambda\, a_i^2}\Big)^2 \,,\label{blodd}
\eea
%%%
where $V$ is defined by
%%%%%
\be
V\equiv \fft1{r^2}\, (1-\lambda\, r^2)\, \prod_{i=1}^n (r^2 + a_i^2)
= \fft{U}{F}\,,
\ee
%%%%%
$W$ and $F$ are given in (\ref{WFdef}), and $U$ is given in (\ref{oddcase3}).
The metric (\ref{blodd}) is obtained from (\ref{kdsodd})--(\ref{oddcase3})
by means of the coordinate transformations
%%%%%
\be
dt=d\tau + \fft{2M\, dr}{(1 - \lambda\, r^2)(V-2M)}\,,\qquad
d\phi_i = d\varphi_i - \lambda\,a_i\,d\tau +
\fft{2M\,a_i\, dr}{(r^2 + a_i^2)(V-2M)}\,.\label{coordtrans}
\ee
%%%%%

   It is useful to note that the sub-determinant for the sector of the
metric involving the $x^\a=(\tau,\varphi_i)$ directions is given by
%%%%%
\be
\det(g_{\a\beta}) = -r^2\, (V-2M)\, W\, 
\prod_{i=1}^n\fft{\mu_i^2}{1+\lambda\, a_i^2}\,.\label{subdetodd}
\ee
%%%%%

   In even spacetime dimensions, $D=2n$, the Kerr-de Sitter metrics
in Boyer-Lindquist form are given by
%%%
\bea
ds^2 &=& - W\, (1 - \lambda r^2)\, 
d\tau^2 + \fft{U\, dr^2}{V-2M} +
\fft{2M}{U}\Bigl(d\tau - \sum_{i=1}^{n-1} \fft{a_i\, \mu_i^2\, d\varphi_i}{
1 + \lambda\, a_i^2}\Bigr)^2 \nn\\
&&+ \sum_{i=1}^n \fft{r^2 + a_i^2}{1 + \lambda\, a_i^2}
\, d\mu_i^2 + \sum_{i=1}^{n-1} \fft{r^2 + a_i^2}{1 + \lambda\, a_i^2}\,
\mu_i^2\, (d\varphi_i-\lambda\, a_i\, d\tau)^2 \nn\\
&&+
\fft{\lambda}{W\, (1-\lambda r^2)}
\Big( \sum_{i=1}^n \fft{(r^2 + a_i^2)\mu_i\, d\mu_i}{
1 + \lambda\, a_i^2}\Big)^2 \,,\label{bleven}
\eea
%%%
where $V$ is defined here by
%%%%%
\be
V\equiv \fft1{r}\, (1-\lambda\, r^2)\, \prod_{i=1}^{n-1} (r^2 + a_i^2)
= \fft{U}{F}\,,\label{Veven}
\ee
%%%%%
$W$ and $F$ are given in (\ref{WFdef}), and $U$ is given in
(\ref{evencase3}).  As usual in the even-dimensional case, there is no
rotation associated with the $i=n$ direction, and so $a_n=0$.  The
metric (\ref{bleven}) is obtained from
(\ref{kdseven})--(\ref{evencase3}) by means of the same coordinate
transformations (\ref{coordtrans}), with the usual understanding that
the azimuthal coordinates arise for the index range $1\le i\le n-1$
only.  The sub-determinant $\det(g_{\a\beta})$ is
%%%%%
\be
\det(g_{\a\beta}) = - r\, (V-2M)\, W\, 
\prod_{i=1}^{n-1}\fft{\mu_i^2}{1+\lambda\, a_i^2}\,,\label{subdeteven}
\ee
%%%%
with $V$ now given by (\ref{Veven}). 

\section{Horizons}

We consider a metric of generalized Boyer-Lindquist form
%%%%%
\ben
ds^2 = X d\tau^2 + 2 Y_i \, d\tau\,  d \varphi ^i + Z_{ij}\, 
d \varphi^i \, d \varphi^j+ g_{ab} dx^a dx ^b\,,
\een
%%%%%
where the angles $\varphi^i$ are periodic, with period $2 \pi$.  The
metric $g_{ab}$ and the quantities $X,Y_i, Z_{ij}$ depend only on the
$x^a$ coordinates $(r,\mu_i)$.  The metric will be free of closed
timelike curves (CTC's), or almost-CTC's, as long as $Z_{ij}$ is
positive definite.

   The surfaces $r=\const$ are invariant under time translations, and
will be timelike as long as the metric $ds^2 = X d\tau^2 + 2 Y_i d\tau
\,d \varphi ^i + Z_{ij} \, d \varphi^i \, d \varphi^j$ is
Lorentzian. To find the intersection of the light-cone with the
surfaces, we set $d\varphi^i =\Omega ^i d\tau$ and $d\mu_i=0$, and
look for null directions.  This leads to an equation for the angular
velocities $ \Omega ^i$:
%%%%%
\ben
Z_{ij}(\Omega ^i + Y^i)(\Omega ^j + Y^j) = Z^{ij} Y_i Y_j -X\,,
\label{cone} 
\een
%%%%%
with $Y^i\equiv Z^{ij}Y_j$ and $Z^{ij} Z_{jk}=\delta ^i_k$.

   Far from the the horizon in the stationary region, $\pa \over \pa
\tau $ is timelike and $X$ is negative. The left-hand side of
(\ref{cone}) is positive, and there is a cone of possible light-like
directions. $X$ may vanish on a timelike ergosurface, but if
$Y_i\ne0$, the right-hand side of (\ref{cone}) will remain positive
and the surfaces $r=\const$ will remain timelike.  As the horizon is
approached, the right-hand side of (\ref{cone}) approaches zero.  The
limiting stationary surface $r=\const$ now contains a single null
direction, and thus it is a stationary null hypersurface.  This is
what Carter \cite{Carter2} calls a Killing horizon.  The null
direction is given by the angular velocities of the horizon $\Omega
^i=\Omega ^i_H$, where
%%%%%
\ben
\Omega_{\sst H} ^i=-Y^i |_{r=r_{\sst H}}\,.
\een
%%%%%

    {\it A priori}, one might imagine that the angular velocities
$\Omega ^i_H$ would depend on the latitudinal coordinates
$\mu_i$. However, the rigidity property of Killing horizons discussed
by Carter \cite{Carter2} implies that they should in fact be {\sl
constant} on the horizon. This is indeed the case in our situation.
In fact, we find
%%%%%
\be
\Omega_{\sst H}^i = \fft{a_i\, (1+\lambda\, a_i^2)}{r_{\sst H}^2 + a_i^2}\,,
\label{Omega}
\ee
%%%%%
where $r_{\sst H}$ is the radius of the horizon.  From
(\ref{subdetodd}) or (\ref{subdeteven}), this occurs at a root of
$V-2M=0$.  It follows that the null generator $l$ of the horizon
coincides with an orbit of the Killing vector field
%%%%%
\ben
l= {\pa \over \pa \tau} +\Omega^i_{\sst H} \, {\pa \over \pa \varphi^i} \,.
\een
%%%%%
In the Kerr-Schild coordinates of section \ref{kerrschildsec}, one has 
%%%%%
\be
\fft{\del}{\del t} = \fft{\del}{\del \tau} + \lambda\,\sum_i\, 
 a_i\, \fft{\del}{\del \varphi_i}\,,\qquad
\fft{\del}{\del \phi_i} = \fft{\del}{\del\varphi_i}\,,
\ee
%%%%% 
and hence
%%%%%
\be
l = \fft{\del}{\del t} + 
\sum_i 
\fft{a_i\, (1-\lambda\, r_{\sst H}^2)}{r_{\sst H}^2+a_i^2}\, 
\fft{\del}{\del \phi_i}\,.
\ee
%%%%%%
Note that the Kerr-Schild vector field $-k^\mu\, \del_\mu$ in section
(\ref{kerrschildsec}) is future-directed and inward-pointing.  In
particular, it crosses the horizon, whilst on the horizon the vector
field $l$ lies in the horizon.  The Kerr-Schild coordinates extend
through the future horizon.  By contrast, the Boyer-Lindquist
coordinates are valid only outside the horizon.

    After some calculation, we find that the area of the horizon is
given by
%%%%%
\bea
D=2n+1:\qquad && A_{\sst H} = \fft{ {\cal A}_{2n-1} }{r_{\sst H}}\, 
  \prod_{i=1}^n 
   \fft{r_{\sst H}^2 + a_i^2}{1+\lambda\, a_i^2}\,,\\
D=2n:\qquad && A_{\sst H} = {\cal A}_{2n-2}\, \prod_{i=1}^{n-1} 
   \fft{r_{\sst H}^2 + a_i^2}{1+\lambda\, a_i^2}\,,
\eea
%%%%%
where
%%%%%
\be
{\cal A}_{m} = \fft{2 \pi^{(m+1)/2}}{
   \Gamma[ (m +1)/2]}\label{snvol}
\ee
%%%%%
is the volume of the unit $m$-sphere.

    Although $l$ is tangent to the null-geodesic generators of the horizon,
it is not affinely parameterised, but rather, 
%%%%%
\be
l^\mu\, \nabla_\mu\, l_\nu = \kappa\, l_\nu\,,
\ee
%%%%%%
where $\kappa$ is the surface gravity, which is constant on each connected
component of the horizon.  Since $l$ is also Killing, we have
%%%%%
\be
\ft12 \nabla_\nu\, L^2 = \kappa\, l_\nu\,,
\ee
%%%%%
where 
%%%%%
\be
-L^2 \equiv l^\mu\, l_\mu = X + 2Y_i\, \Omega_{\sst H}^i + Z_{ij}\, 
\Omega_{\sst H}^i\, 
     \Omega_{\sst H}^j\,.
\ee
%%%%%
Thus we have
%%%%%
\be
\kappa^2=(\nabla_\mu L)\, (\nabla^\mu L) \,.
\ee
%%%%%
Note that $L$ vanishes on the horizon, but $\nabla_\mu\, L$ is non-zero.
Near the horizon we find that
%%%%%
\be
L^2 \approx 
  U(r_{\sst H})\, \Bigl( \fft{1-\lambda\, 
r_{\sst H}^2}{V(r_{\sst H})}\Bigr)^2\,
   V'(r_{\sst H})\, (r-r_{\sst H})\,,
\ee
%%%%%
and hence we find 
%%%%%%
\be
\kappa = \ft12 (1-\lambda\, r_{\sst H}^2)\, 
\fft{V'(r_{\sst H})}{V(r_{\sst H})}\,.\label{kappavp}
\ee
%%%%%
In odd and even dimensions, this gives
%%%%%
\bea
D=2n+1:\qquad && \kappa= r_{\sst H}(1-\lambda\, r_{\sst H}^2)\, \sum_{i=1}^n 
\fft1{r_{\sst H}^2 + a_i^2} - \fft1{r_{\sst H}}\,,\label{kappaodd}\\
D=2n:\qquad && \kappa= r_{\sst H}(1-\lambda\, r_{\sst H}^2)\, \sum_{i=1}^{n-1} 
\fft1{r_{\sst H}^2 + a_i^2} - \fft{1+\lambda\, 
r_{\sst H}^2}{2r_{\sst H}}\,.\label{kappaeven}
\eea
%%%%%

\section{Compact Euclidean-Signature Einstein Metrics}

    In this section, we consider the Euclidean-signature metrics
(positive-definite-signature, Riemannian as opposed to
pseudo-Riemannian or Lorentz-signature) that are obtained from the
Boyer-Lindquist forms (\ref{blodd}) and (\ref{bleven}) of the Kerr-de
Sitter metrics, by making the replacements
%%%%%
\be
\tau = -\im\, \psi\,,\qquad a_i = \im\, \a_i\,.
\ee
%%%%%
Under these transformations the metrics are again real (with real
values for $\psi$ and $\a_i$), and now have positive-definite
signature.

   For general values of the rotation parameters $\a_i$ and mass
parameter $M$, the Euclidean-signature metrics do not extend smoothly
onto complete, compact, non-singular manifolds.  However, this can be
achieved for special choices of the parameters, as we shall now show.
The discussion follows the one given in \cite{pagekerr} for the
four-dimensional Euclidean-signature Kerr-de Sitter metric, and in
\cite{hassakyas} for the five-dimensional Euclidean-signature Kerr-de
Sitter metric.

   So far, we have presented the Kerr-de Sitter metrics in a local
coordinate system, and the ranges of the coordinates $r$, $\psi$ and
$\varphi_i$ are {\it a priori} undetermined.  The metrics have various
coordinate singularities, at which either components of the metric
diverge, or the determinant vanishes.  Geometrically, the coordinate
singularities arise because the $U(1)^{N+1}$ isometry group generated
by $\del/\del\varphi^i$ and $\del/\del\psi$ has fixed points.  In the
general case, there is a single linear combination of
$\del/\del\varphi^i$ and $\del/\del\psi$ that vanishes at each
fixed-point set.  If the manifold is to be smooth and non-singular,
and the $U(1)^{N+1}$ isometry group is to act smoothly, then the
fixed-point sets must be non-singular embedded sub-manifolds.

   From equations (\ref{subdetodd}) or (\ref{subdeteven}), the
fixed-point sets occur where any of the $\mu_i$ coordinates vanishes,
and where $V=2M$.  Near the fixed-point sets, we introduce local
coordinates and determine the periods of $\psi$ and $\varphi_i$ that
permit the elimination of the local coordinate singularities. We then
must consider the compatibility conditions between these different
local coordinate charts.  These compatibility conditions give rise to
non-trivial constraints on the parameters in the metric.

    The simplest metric singularities to consider are those associated
with the vanishing of the $\mu_i$ coordinates.  As in flat space,
regularity requires that each coordinate $\varphi_i$ be periodic
with period $2\pi$, keeping the remaining coordinates fixed.

   Further coordinate singularities occur where 
$V-2M=0$. We begin by considering one such root, situated say at $r=r_1$.
Near $r=r_1$ the metric takes the form
%%%%%
\be
ds^2 \approx d \rho ^2 + \kappa _1^2 \, \rho^2 \, d \psi^2 + g_{ij}\,
 ( d \varphi^i -\Omega^i_1 \, d \psi ) ( d \varphi^j - \Omega ^j_1\,  
d \psi ) + \tilde g_{ij} \, d \mu^i d \mu^j\,, 
\ee
%%%%%
where $\rho $ is the distance from the horizon and $\Omega_1^i$ is now
the Euclidean angular velocity, defined to be 
$-\im\, \Omega_{\sst H}^i$ given by (\ref{Omega}) with $a_i=\im\, \a_i$, 
namely 
%%%%%
\be
\Omega_1^i = \fft{\a_i\, (1-\lambda\, \a_i^2)}{r_1^2 -\a_i^2}\,.
\ee
%%%%%
 The conical singularity  at $\rho =0$ 
can be removed by the introduction of local Riemann normal coordinates
provided that the period of $\psi$ is chosen to be 
$2\pi/\kappa_1$ at fixed 
%%%%%
\be
\varphi_1^i= \varphi ^i-\Omega ^i_1 \psi.
\ee
%%%%%
For future use we perform a further local coordinate transformation to
$\psi_1= \kappa_1\, \psi$, which has period $2 \pi$.

   If the cosmological constant is positive, any smooth solution must
be compact.  If the cosmological constant is negative, then there does
not exist any smooth compact solution with continuous isometries. Thus
there are essentially only two possibilities to consider: either
$\lambda$ is positive and we have a compact smooth solution for which
the range of $r$ lies between two roots of $V-2M=0$, at $r=r_1$ and
$r=r_2$, or $\lambda$ is negative and hence the solution is not
compact.  We shall consider only the former case.

    We have two local charts with coordinates $(\psi_1, \varphi_1^i)$
and $(\psi_2, \varphi^i_2)$, all of which have period $2\pi$.  The
transition function between them is given by the matrix $S$ effecting
the linear coordinate transformation
%%%%%
\be
\pmatrix{\psi_2\cr \varphi^i_2} = \pmatrix{\fft{\kappa_2}{\kappa_1} & 0\cr
       \fft{\Omega^i_1-\Omega^i_2}{\kappa_1} & \delta^i_j}\,
\pmatrix{\psi_1 \cr \varphi^j_1}\,.
\ee
%%%%%
The transition function is a map between two parameterisations of the
torus $T^{N+1}$, where $N=[(D-1)/2]$.  It will be well-defined and
invertible if and only if the matrix $S$ is invertible, and both it
and its inverse have integer entries.  In other words, $S$ is an
element of $SL(N+1,\Z)$.  This leads immediately to the requirements
%%%%%
\bea
|\kappa_1| = |\kappa_2| &\equiv& \kappa\,,\\
\Omega_1^i - \Omega_2^i &=& \kappa\, k_i\,,\label{kidef}
\eea
%%%%%
where the $k_i$ are integers.  If we take $r_1 < r_2$, then we shall
have $\kappa_1=-\kappa_2 =\kappa > 0$.   

   The result of this construction is a $T^{N}$ bundle over $S^2$.
Such bundles are given by a map from the equator of $S^2$ to $T^{N}$,
specified by the integers $k_i$, which characterise the windings of
the image of the equator around the $N$ cycles of the torus.  The
manifold as a whole is an associated $S^{D-2}$ bundle over $S^2$, with
structure group $T^{N}$, whose action on $S^{D-2}$ is just rotations
around the azimuthal coordinates $\varphi^i$.  The bundle is trivial
or non-trivial according to whether $\sum_i k_i$ is even or odd.

   Equation (\ref{kidef}) implies
%%%%%
\be
\fft{\a_i\, (1-\lambda\,\a_i^2)\, (r_2^2-r_1^2)}{\kappa\, (r_1^2-\a_i^2)
(r_2^2-\a_2^2)} = k_i\,.\label{kidef2}
\ee
%%%%%%
The equality of $|\kappa_1|$ and $|\kappa_2|$ can be achieved by
allowing the roots at $r=r_1$ and $r=r_2$ to coincide.  (Of course,
the distance between the roots in this limit is non-vanishing, since
$g_{rr}$ diverges.)  It is convenient to write $r_1=r_0-\ep$ and
$r_2=r_0+\ep$, and take the limit $\ep\rightarrow 0$.  In this limit,
the radius $r=r_0$ therefore corresponds to a double root of $V-2M$.
The numerator and denominator of (\ref{kidef2}) both go to zero, with
the ratio giving the finite limit
%%%%%
\be
  k_i = - \fft{8 r_0\, V(r_0)\, \a_i\, (1-\lambda\, \a_i^2)}{
  (1-\lambda\, r_0^2)\,
 V''(r_0)\, (r_0^2-\a_i^2)^2}\,.\label{kidef3}
\ee
%%%%%

   It is convenient to define the dimensionless parameters 
%%%%%
\be
\nu_i \equiv \fft{\a_i}{r_0}\,,
\ee
%%%%%
and the quantities
%%%%%
%\be
%B \equiv  \sum_{i=1}^N \fft{\nu_i^2}{(1-\nu_i^2)^2}\,,\qquad
%C \equiv \sum_{i=1}^N \fft{\nu_i^2}{1-\nu_i^2}\,,\nn
%\ee
\bea
\a &\equiv & \sum_{i=1}^N \fft{1}{1-\nu_i^2}
+ \sum_{i=1}^n \fft{1}{1-\nu_i^2} 
\,=\, D-1+2\sum_{i=1}^N \fft{\nu_i^2}{1-\nu_i^2}\,, \nn \\
\beta &\equiv &  \fft{\a}{\a-2} \,, \nn \\
\gamma &\equiv &
1 + \fft{4\b}{\a^2}\,\sum_{i=1}^N \fft{\nu_i^2}{(1-\nu_i^2)^2} \, ,
\label{abcdef}
\eea
%%%%%
where, as usual, $N=[(D-1)/2]$ is the number of azimuthal coordinates
$\varphi^i$ (and also the number of angular momentum parameters
$\a_i$), and $n=[D/2]$ is the number of latitudinal coordinates
$\mu_i$, which obey the constraint $\sum_i\mu_i^2=1$ (so that only
$n-1$ of them are independent), which along with the coordinates $r$
and $\psi$ gives the total spacetime dimension $D=N+n+1$.  Note that
for odd $D=2n+1=2N+1$, we have $n=N$, but for even $D=2N+2=2n$, we
have $n=N+1$ (and then $\nu_n=0$).

   For $r_0$ to be a double root of $V-2M=0$, we need to set
$2M=V(r_0)$ and $V'(r_0)=0$.  From (\ref{kappavp}), the latter
condition implies that $\kappa(r_0)=0$.  For all dimensions, odd or
even, the conditions for the double root can be solved to give
%%%%%
\be
r_0 = (\b \lambda)^{-1/2} \,,\qquad
M= \a^{-1}\, r_0^{D-3}\, \prod_{i=1}^N (1-\nu_i^2)\,.
\ee
%%%%%

   Substituting into (\ref{kidef3}) then gives
%%%%%
\be
k_i = \fft{4\nu_i\, (\b\,-\, \nu_i^2)}
      {\a\g\, (1-\nu_i^2)^2\,}\,,
\label{regular}
\ee
%%%%%
which are the non-trivial conditions that must be satisfied in order
for the metrics to be regular.

   In order to obtain the metrics on the $S^{D-2}$ bundles over $S^2$,
we define a new radial coordinate $\chi$ by writing 
%%%%%
\be
r = r_0 -\ep\, \cos\chi
\ee
%%%%%
prior to taking the limit $\ep\rightarrow 0$.  The metric
in the $\ep\rightarrow 0$ limit becomes

%%%%%
\bea
\lambda ds^2 &=& \fft{Z}{\a\gamma}\, 
   (d\chi^2 + \sin^2\chi\, d\psi_1^2) +  
\sum_{i=1}^n\, \fft{1-\nu_i^2}{\b-\nu_i^2}\, d\mu_i^2 
  + \fft{\a}{2 W}\, \Big( \sum_{i=1}^n
 \fft{1-\nu_i^2}{\b-\nu_i^2} \, \mu_i\, d\mu_i \Big)^2\, \nn \\
&&
 + \sum_{i=1}^N \fft{1-\nu_i^2}{\b-\nu_i^2} \, \mu_i^2\, 
    (d\varphi_1^i + k_i\, \sin^2\fft{\chi}{2}\, d\psi_1)^2 \nn\\
&&
 - \fft{2\b}{\a Z}\, \Big[\sum_{i=1}^N \fft{\nu_i\, \mu_i^2}{\b-\nu_i^2}\,
            (d\varphi_1^i + k_i\, \sin^2\fft{\chi}{2}\,
  d\psi_1)\Big]^2 \,,
\label{eumetric}
\eea 
%%%%%
where $\a$, $\b$, and $\g$ are the $\nu_i$-dependent constants
given by (\ref{abcdef}), $W$ and $Z$ are functions of the latitudinal
coordinates $\mu_i$ given by 
\be
W \equiv \sum_{i=1}^n \fft{\b\, \mu_i^2}{\b-\nu_i^2}
 = 1 + \sum_{i=1}^N \fft{ \nu_i^2\, \mu_i^2}{\b-\nu_i^2} \, ,
\label{Wdef}
\ee
\be
Z \equiv \sum_{i=1}^n \fft{\mu_i^2}{1-\nu_i^2}
 = 1 + \sum_{i=1}^N \fft{\nu_i^2\,
\mu_i^2}{1-\nu_i^2}\,,
\label{Zdef}
\ee
and the constants $k_i$ are given by
(\ref{regular}) (and must be integers for regularity at $\chi = \pi$,
thus permitting only discrete choices for the $\nu_i$'s).

   If the metric (\ref{eumetric}) has positive-definite signature with
eigenvalues that are neither zero nor infinite for $0 < \chi < \pi$
and for $0 < \mu_i^2 < 1$, then $W$, $Z$, and $\a$ must be finite,
nonzero, and have the same sign for all allowed values of $\mu_i$.
Since one can set any one of the $\mu_i^2$ arbitrarily close to 1 by
having the others arbitrarily close to 0, one can see from
(\ref{Wdef}) and (\ref{Zdef}) that it is both necessary and sufficient
that all the values of $1-\nu_i^2$ and of $\b-\nu_i^2$ for all $i$
must be non-zero and all have the same sign.  We exclude and leave for
later investigation the special cases in which one or more of the
$\nu_i^2$'s are unity.\footnote{For example, there is a special case
in $D=5$ with $(k_1,k_2)=(0,2)$ and $\nu_1=1$, which corresponds to
the round metric on $S^5$ \cite{hassakyas}.}  The discussion that
follows divides into two cases, according to whether all $\nu_i$
satisfy $0\le \nu_i^2 <1$, or all satisfy $\nu_i^2>1$.

\bigskip
\noindent{\underline{{\bf $0\le \nu_i^2 <1$ for all ${\nu_i}$}} }\,\,:
\medskip

   If $D$ is even, $D = 2n = 2N+2$, we have $\nu_n=0$, and thus
 regularity requires $\nu_i^2 < 1$ for all $i$.  This is also required
 for $D$ odd, $D = 2n+1$, if one or more of the $k_i$'s (and hence the
 corresponding $\nu_i$'s) is zero.  More generally, if $\nu_i^2 < 1$,
 it follows from (\ref{abcdef}) that $\a \ge D-1$, and $1 < \b \le
 (D-1)/(D-3) \le 3$.  The quantity $\b$ attains its maximum value of
 $(D-1)/(D-3)$ if and only if all $\nu_i = 0$, which leads to the
 product metric on $S^2 \times S^{D-2}$.  Since $\b > 0$, the second
 term of the right-hand side of the expression for $\g$ in
 (\ref{abcdef}) is non-negative, so $1 \le \g$, with equality only for
 the product metric on $S^2 \times S^{D-2}$, for which all $\nu_i =
 0$.

   We have that 
%%%%%
\bea
\fft{\a^2}{\b}
&=& \a(\a-2)\, \ge \, 4\left(\sum_{i=1}^N \fft{1}{1-\nu_i^2}\right)^2
-4\sum_{i=1}^N \fft{1}{1-\nu_i^2} \nn \\
&>& 4\sum_{i=1}^N \fft{1}{(1-\nu_i^2)^2}
-4\sum_{i=1}^N\fft{1}{1-\nu_i^2}
\, = \, 4\sum_{i=1}^N \fft{\nu_i^2}{(1-\nu_i^2)^2} \, ,
\label{ineq}
\eea 
%%%%%
and therefore  from (\ref{abcdef}) we get $1 \le \g < 2$.

   Since $\b > 1$, $\b-\nu_i^2$  and  $1-\nu_i^2 > 0$ have the same sign, 
and it follows that  $W$, $Z$, and $\a$ all have the
same sign for all allowed values of the $\mu_i$'s.  This  gives  a
positive-definite nonsingular metric (\ref{eumetric}) if the
regularity condition (\ref{regular}) is satisfied.
 
   We now show that if $\nu_i^2 < 1$ for all $i$,
the regularity condition (\ref{regular}) can be satisfied only 
if either the metric is a product with  $\nu_i = 0$, or if just
 one $k_i$ is nonzero and takes the value $\pm 1$.
For simplicity of  presentation, we shall reverse the sign of any
coordinate $\varphi_1^i$ corresponding to a negative $\nu_i$, so that,
without loss of generality, we  have all $\nu_i$ non-negative, and hence
all $k_i$ non-negative.

   If there is at least one positive $\nu_i$, say of value $\nu$
with $0 < \nu < 1$, then (\ref{abcdef}) implies that 
%%%%%
\be
\a \ge D-3+\fft{2}{1-\nu^2}\, .
\ee
%%%%%
Using the fact that $\nu_i^2 < 1$, we have $\b > 1$ and
$\g > 1$ . This is now  a strict inequality, since the sum in the formula
(\ref{abcdef}) for $\g$ is positive with at least the one positive
$\nu_i$.  The regularity condition (\ref{regular}) for the $k_i$
corresponding to the positive $\nu_i=\nu$ becomes
\be
k_i = \fft{4\nu\, (1\,-\, \nu_i^2/\b)}
      {(\a-2)\g\, (1-\nu_i^2)^2\,} < \fft{4\nu}{(\a-2)\g\,
        (1-\nu_i^2)}
    < \fft{4\nu}{2+(D-5)(1-\nu^2)} < 2\,
\label{reg1}
\ee
for $D \ge 5$.  For $D=4$, it was already proved in \cite{pagekerr}
that $k_i$, there called $n$, must be less than 2.  Therefore, when
$\nu_i^2 < 1$, all $k_i$ must be either 0 or 1, given that  we have chosen
coordinates so that all $k_i$ are non-negative.

   Next, we prove that there cannot be more than one positive
integer $k_i$, by assuming the contrary  and obtaining a contradiction.
From (\ref{regular}) at  fixed $\a$, $\b$, and
$\g$, and  using the facts that   $0 \le \nu_i < 1$ and hence  $\b > 1$,
it follows   that $dk_i/d\nu_i > 0$. Thus,   for given  $k_i$, there is only
one $\nu_i$ in the allowed range that solves the  regularity
condition. By the inequality deduced above, 
the nonzero $k_i$'s must  all equal $1$.  Therefore all  nonzero
$\nu_i$'s must  take  the same value, say $\nu$.

   If we assume that there are at least two nonzero $\nu_i$'s that
have the same  positive value $\nu$, we get 
\be
\a \ge D-5+\fft{4}{1-\nu^2}\, ,
\ee
and  the final two inequalities of (\ref{reg1}) become
\be
k_i < \fft{4\nu}{4+(D-7)(1-\nu^2)} < 1\,
\label{reg2}
\ee
for $D \ge 7$. Thus the $k_i$ must vanish.

   The two remaining cases for which one might have two nonzero
$\nu_i$'s are $D=5$ and $D=6$. Since both of these cases give $N=2$,
only two non-zero $\nu$'s are allowed in each of those two
cases.  These require special arguments to show that $k_i < 1$.  These
arguments run as follows.  For $D=5$ and $\nu_1 = \nu_2 = \nu$, we get
$\a = 4/(1-\nu^2)$, $\b = 2/(1+\nu^2)$, and $\g =
(1+2\nu^2)/(1+\nu^2)$. Thus $k_i = \nu(2+\nu^2)/(1+2\nu^2)$, which one
can easily show is less than unity for $0 \le \nu < 1$.  For $D=6$ and
$\nu_1 = \nu_2 = \nu$, we get $\a = (5-\nu^2)/(1-\nu^2)$, $\b =
(5-\nu^2)/(3+\nu^2)$, and $\g = (15+10\nu^2-\nu^4)/(15+2\nu^2-\nu^4)$
Thus $k_i = 4\nu(5+\nu^2)/(15+10\nu^2-\nu^4)$, which one can also
easily show is less than unity for $0 \le \nu < 1$.  Therefore, for
all $D \ge 5$, for which more than one nonzero $\nu_i$ is allowed by
the local solution, we find that one cannot actually have more than
one nonzero $\nu_i$ for a regular complete compact Einstein metric,
assuming $\nu_i^2 < 1$.

   To summarize: choosing coordinates such that $k_i\ge 0$ and hence
$\nu_i\ge0$, we find that for $\nu_i < 1$ we can either have all
$\nu_i = 0$, or we have one and only one $\nu_i$ nonzero, and hence
just one non-zero $k_i$, which must take the value $1$.  This means
that even though our local solution allows up to $N$ nonzero rotation
parameters, for complete compact Einstein metrics of Euclidean
(positive-definite) signature with any $\nu_i^2 < 1$, there can be
only one nonzero rotation parameter, and our solutions reduce to those
of \cite{hassakyas}.  The condition $\nu_i^2 <1$ is necessarily the
case when $D$ is even, and possibly the case when $D$ is odd.

\bigskip
\noindent{\underline{{\bf $\nu_i^2>1$ for all $\nu_i$}}}\,\,:
\medskip

   As discussed previously, if the dimension $D=2N+1$ is odd, we can
also have regular metrics in which all $\nu_i^2>1$.  This requires
that all the integers $k_i$ be non-zero. In fact, we can show that
there are infinitely many inequivalent smooth complete compact
Einstein metrics for each odd dimension.  In general, there appears to
be a unique such metric for each possible choice of the $N$ positive
integers $k_i$, although for $k_1=k_2=1$ in $D=5$ one has the special
case $\nu_1=\nu_2=1$, which is outside the scope of the present
discussion.  (It leads to a regular compact metric, namely the
homogeneous Einstein metric on $T^{1,1}$ \cite{hassakyas}.)

   If all  $\nu_i$ satisfy $\nu_i^2>1$, one has  from (\ref{abcdef})
that $\a < 0$, and hence $0 < \b < 1$ and $1 < \g$. 
Using the fact that all the $1-\nu_i^2$ terms have the same sign, so
that the cross terms in the square of the sum contribute positively,
we again  get from (\ref{ineq})  that $\g < 2$.
Thus, in this case the allowed range for $\g$ is
$1 < \g < 2$.  Because in this case  $\b < 1$,  a  sufficient
condition for  a
positive-definite nonsingular metric (\ref{eumetric}) satisfying the
regularity condition (\ref{regular}) for odd $D=2N+1$  is 
that both $1-\nu_i^2$ and $\b-\nu_i^2$ have
the same negative  sign for all $i$. 

  For this $\nu_i^2>1$ case, it is useful to define
%%%%%
\bea
&&x_i \equiv \fft{1}{\nu_i^2-1} \,,\qquad
A   \equiv \sum_{i=1}^N x_i \,,\qquad
B  \equiv \sum_{i=1}^N x_i^2 \, , \nn \\
&&C  \equiv \fft{1}{A^2+2A+B} \,, \qquad
y_i \equiv  A+1+x_i \, .
\label{ABCdef}
\eea
%%%%%
Then $\a = -2A$, $\b = A/(A+1)$, $\g = 1/[A(A+1)C]$, and the
regularity condition (\ref{regular}) becomes
\be
k_i = 2\, C \, y_i \, \sqrt{x_i(1+x_i)}\, .
\label{reg3}
\ee

   Each $x_i$ monotonically decreases from $x_i=\infty$ for $\nu_i=1$
to $x_i=0$ for $\nu_i=\infty$, and thus prior to imposing the
regularity condition (\ref{reg3}) that each $k_i$ be a positive
integer, each $x_i$ would be allowed to range over the entire positive
real axis.

   Now define
%%%%%
\be
M_{ij} \equiv \fft{\partial \log{k_i}}{\partial x_j}
= D_i \delta_{ij} + a_i + b_j \, ,
\ee
%%%%%
where
%%%%%
\be
D_i = \fft{1}{y_i} + \fft{1}{2x_i} + \fft{1}{2(1+x_i)} \,,\qquad
a_i = -2C y_i \,,\qquad
b_j = \fft{1}{y_j} \, .
\ee
%%%%%
   The Jacobian of the transformation from the $x_i$ to the
$\log{k_i}$ is the determinant of the matrix $M_{ij}$, which, it
appears, is always negative for all positive $x_i$.\footnote{We have
verified this explicitly for all $N \le 5$, and numerically in
numerous higher-dimensional cases.}  Therefore, the map from all
positive $x_i$ to $\log{k_i}$ (and hence to the $k_i$) is
surjective. Thus if there is any solution for the set of $x_i$ given a
set of positive $k_i$, this solution must be unique. 

   Extensive numerical searches strongly suggest that for almost all
sets of positive integers $k_i$ ($k_1=k_2=1$ for $D=5$ being the only
exception we have found, though as noted above it also leads to a
regular compact metric, on $T^{1,1}$ \cite{hassakyas}), there is a
corresponding unique finite positive solution for the $x_i$'s and
hence for the $\nu_i=\sqrt{(1+x_i)/x_i}$.

   It follows that for each set of $N$ entirely positive integers
 $k_i$, there is a corresponding smooth complete compact Einstein
 metric in $D=2N+1 \ge 5$ dimensions.  This means that there are
 infinitely many smooth Einstein metrics in all odd dimensions $D\ge
 5$.  In particular, since the case in which $\sum_i k_i$ is even
 gives the trivial $S^{D-2}$ bundle over $S^2$, we have infinitely
 many smooth Einstein metrics on $S^2 \times S^{D-2}$ for all odd
 $D\ge 5$.
 
    It is worth remarking that the $\nu_i$ are given approximately by
%%%%%
\be
\nu_i \approx k_i^{-1}\, \sum_{j=1}^N k_j^2
\ee
%%%%%
in the case that $\sum_{j=1}^N k_j^2 \gg 1$, which applies for all but a
small finite number of cases.  In particular,   
this approximation is a good one for any $N$ if at least one of 
the $k_i$ is large, and it is also a good approximation for arbitrary
positive $k_i\ge1$ if $N$ is large.

     An better approximation that is always accurate within $0.026\%$
error is
%%%%%%
\bea
\nu_i \approx \fft{1}{a k_i}
 \
 \Big[\!\!\!\!\!\!\!\!\!\!\!&&1-\fft{1}{2}a(1+5b-4\lambda_i)
 -\fft{1}{2}a^2(b+25b^2-24c-10b\lambda_i+10\lambda_i^2)  \nn \\
 &&+\fft{1}{8}a^3(1+42b-43\lambda_i+167b^2-216c-252b\lambda_i+304\lambda_i^2
 \nn \\
 &&-192b^3+233bc-521b^2\lambda_i+497c\lambda_i+486b\lambda_i^2-529\lambda_i^3)
 \Big] \, , \label{approx}
\eea
%%%%%%
where
%%%%%
\be
a \equiv \left(\sum_{i=1}^N k_i^2\right)^{-1} \,,\qquad
\lambda_i \equiv a k_i^2 \,,\qquad
b\equiv \sum_{i=1}^N \lambda_i^2\,,\qquad 
c\equiv \sum_{i=1}^N \lambda_i^3\,.
\ee
%%%%%

     The terms inside the square brackets that are proportional to
unity, $a$, and $a^2$ are the first three terms in the power series
expansion in $a$ for small $a$.  The terms proportional to $a^3$ do
not form the next term in the power series in $a$ but were chosen to
make the error small for cases in which $a$ is not small.

     In particular, four linear combinations of the thirteen integer
coefficients terms inside the round brackets that form the coefficient
of $\fft{1}{8}a^3$ were determined by requiring that when all the $k_i$ are
equal, one gets the first two terms correct (but not the next two) in
an expansion in $1/N$ for the coefficient of $a^3$, that one gets the
correct answer of $\nu_1=\nu_2=1$ when $N=2$ and $k_1=k_2=1$, and that
one gets nearly the correct answer of $\nu_1=\nu_2=\nu_3=1+2^{1/3}$
when $N=3$ and $k_1=k_2=k_3=1$ (as close as possible with integer
coefficients inside the round bracket).  The remaining nine of the
thirteen coefficients inside the round bracket were then chosen to be
the integers closest to the values of the coefficients that minimized
the sum of the squares of the relative errors of the eighteen
different $\nu_i$'s for the nine sets of $k_i$'s that have not all
$k_i$'s equal in each set and which give $a \ge 0.1$.

     This is a somewhat {\it ad hoc} procedure for determining the
form of Eq. (\ref{approx}), but it does lead to an excellent
approximation that has a relative error of more than $10^{-4}$ in only
a handful of cases.  The worse case is that of $N=2$, $k_1=k_2=2$, for
which the correct answer is $\nu_1=\nu_2 = \nu = (4+10^{1/3}+10^{2/3})/3
\approx 3.598\,674\,508$, whereas Eq. (\ref{approx}) gives
$\nu_1=\nu_2=\nu \approx 1843/512 = 3.599\,609\,375 \approx
\nu(1+0.000\,259\,781)$, with a relative error of less than one part
in 3800.

   One can get exact explicit solutions for the $\nu_i$ in some simple
cases. One example for each odd $D=2N+1 \ge 5$ is the case in which
all $N$ values of $k_i$ are equal to $k$ say, and therefore all $N$
values of $\nu_i$ are also equal, say to $\nu$.  The regularity
condition (\ref{regular}) then becomes
%%%%%
\be
k = \fft{2\nu(N+\nu^2)}{N(N-1+2\nu^2)} \, .
\ee
%%%%%
If one sets $\nu = Nky$, one gets the cubic equation
%%%%%
\be
y^3-y^2+\fft{1}{Nk^2}\left(y-\fft{1}{2}+\fft{1}{2N}\right)=0 \, .
\ee
%%%%%
For large $N k^2$, the solution for $y$ is just a bit smaller than
unity.

   Indeed, solving the cubic explicitly gives
%%%%%
\bea
\nu &=& \fft{1}{3} N k \Big\{ 1+ 
 \Big[1+\fft{9(N-3)}{4N^2 k^2}
  +3^{3/2} \left( \fft{N-1}{2 N^2 k^2} - \fft{13 N^2 + 18 N - 27}{16 N^4 k^4}
          +\fft{1}{N^3 k^6} \right)^{1/2} \Big]^{1/3} \nn \\
&+& \Big[1+\fft{9(N-3)}{4 N^2 k^2}
  -3^{3/2} \left( \fft{N-1}{2 N^2 k^2} - \fft{13 N^2 + 18 N - 27}{16 N^4 k^4}
          +\fft{1}{N^3k^6} \right)^{1/2} \Big]^{1/3} \Big\}\,,
\eea
%%%%%
which has the following series expansion in $a=1/(N k^2)$:
%%%%%
\be
\nu = N k \Big[1-\fft{1}{2}a(1+N^{-1})-\fft{1}{2}a^2(N^{-1}+N^{-2})
+\fft{1}{8}a^3(1-N^{-1}-9N^{-2}-7N^{-3})+O(a^4)\Big] \, . 
\label{expansion}
\ee 
%%%%%

     The first three terms of this series expansion (\ref{expansion})
in $a$ coincide with the corresponding ones of (\ref{approx}) when all
the $k_i$ are equal, but in the $\fft{1}{8}a^3$ term, the coefficients
of $N^{-2}$ and of $N^{-3}$ in the series expansion above, $-9$ and
$-7$ respectively, disagree with what (\ref{approx}) gives for those
coefficients, $+3$ and $-26$ respectively.  The difference is due to
the fact that the series expansion (\ref{expansion}) is that for small
$a$, whereas (\ref{approx}) was designed to fit the results for all
allowed $a$ to high accuracy.

   Special cases of our new compact metrics reduce to previously-known
examples.  When $D=4$, there is only one parameter $\nu_i$, which is
the parameter $\nu$ in \cite{pagekerr}, and (\ref{regular}) reduces to
equation (23) of \cite{pagekerr}.  In arbitrary dimensions, but with
only one $\nu_i$ non-zero, the conditions (\ref{regular}) reduce to
equation (3.22) of \cite{hassakyas}.  In $D=5$, with both parameters
$\nu_1$ and $\nu_2$ non-zero, (\ref{regular}) reduces to equation
(2.22) of \cite{hassakyas}.

    To the best of our knowledge, the metrics (\ref{eumetric}) in
arbitrarily high odd dimensions $D=2N+1$ and with arbitrarily many
different $\nu_i$'s are the first explicit compact Euclidean Einstein
metrics with arbitrarily high cohomogeneity (cohomogeneity $N-1 =
[(D-2)/2]$) that are not merely product metrics or warped products
with a bounded number of warping parameters.

    It is useful to note that the area of the fixed-point set of 
$\del/\del\psi_1$, which occurs at $\chi=0$, is given by
%%%%%
\be
A_{\sst H} = {\cal A}_{D-2} \, 
(\b\lambda)^{-(D-2)/2}\,
\prod_{i=1}^N\, \fft{1 - \nu_i^2}{1 - \nu_i^2/\b}\,,
\label{fps1}
\ee
%%%%%
where ${\cal A}_N$ is the volume of the unit $N$-sphere, given by 
(\ref{snvol}).  The area of the fixed-point set of $\del/\del\psi_2$, 
which occurs at $\chi=\pi$, is also given by (\ref{fps1}).
The volume of the complete compact metric is given by
%%%%%
\be
V_C = {\cal A}_D \, (\b\lambda)^{-D/2} \,
\fft{2\b}{\g} \prod_{i=1}^N\, \fft{1 - \nu_i^2}{1 - \nu_i^2/\b}\,
= \fft{2{\cal A}_{D}}{\g {\cal A}_{D-2}}\, \l^{-1} A_{\sst H} \,.
\ee
%%%%% 
Note that because $1 \le \g < 2$, we have the inequality
${\cal A}_{D} \, A_{\sst H} \, < \, {\cal A}_{D-2} \, \lambda V_C
\le 2{\cal A}_{D}\, A_{\sst H} \, $.

    To summarise, we have seen that complete and non-singular compact
Einstein metrics arise as follows.  In even dimensions $D\ge4$, the
only regular solutions occur when all $k_i$ except one vanish, and the
sole non-vanishing $k_i$ is equal to 1 (or, equivalently modulo
trivial coordinate transformations, $-1$).  Correspondingly, all the
$\nu_i$ vanish except for the one associated with the non-vanishing
$k_i$.  The choice of which of the $k_i$ is the non-vanishing one is
inessential, since all choices are equivalent up to permutation.

   In odd dimensions $D\ge5$ there is an analogous such solution in
which all except one of the $k_i$ vanish, and again all the $\nu_i$
except the one associated with the non-zero $k_i$ vanish.  In all the
above cases, the non-vanishing $\nu_i$ satisfies $0<\nu_i^2<1$.
However, in the odd-dimensional case there are also infinitely many
additional regular metrics, in general one for each choice of positive
integers $k_i$. In these cases, all the $\nu_i$ satisfy
$\nu_i^2>1$.  (We know of one exceptional case, in $D=5$, where 
there is a regular solution with $k_1=k_2=1$, for which $\nu_1=\nu_2=1$;
this gives the homogeneous Einstein metric on $T^{1,1}$ \cite{hassakyas}.)

\section*{Acknowledgments}
  
    We are grateful to Tekin Dereli, Sean Hartnoll and Yukinori Yasui
for useful discussions.  G.W.G. and D.N.P. are grateful to the George
P. \& Cynthia W. Mitchell Institute for Fundamental Physics for
hospitality during the course of this work.

\newpage
\appendix
\bigskip\bigskip
\centerline{\Large {\bf Appendices}}
\bigskip

    In the appendices, we present some special cases of the Kerr-de
Sitter metrics that we have obtained in this paper.  First, in
Appendix A, we give the Kerr-Schild form of the previously-known
Kerr-de Sitter metrics \cite{Carter,hawhuntay} in $D=4$ and $D=5$. Next,
in Appendices B and C, we give the Kerr-de Sitter metrics in all odd
dimensions $D=2n+1$ and even dimensions $D=2n$, in the special case
where the rotation parameters $a_i$ are all set equal. In
Appendix D, we give the special case where all except one of the
rotation parameters are set to zero.  Under this specialisation, our
new metrics reduce to ones that were obtained in arbitrary dimensions
in \cite{hawhuntay}.  Finally, in Appendix E, we give a unified
alternative Boyer-Lindquist form of the metrics that applies to both
even and odd dimensions and also eliminates one set of off-diagonal
terms that appear in the historically-standard Boyer-Lindquist form
with a cosmological constant.

\section{Kerr-Schild Form of $D=4$ and $D=5$ Kerr-de Sitter}

Here, we present the explicit Kerr-Schild form of the Kerr-de Sitter
metrics in dimensions 4 and 5. 

   In $D=4$, we write the coordinates $\mu_i$ in
(\ref{kdseven})--(\ref{evencase3}) as $\mu_1=\sin\theta$,
$\mu_2=\cos\theta$, and we define $\phi_1=\phi$, $a_1=a$.  The Kerr-de
Sitter metric is then written as $ds^2 = d\bar s^2 +\fft{2M}{U}\,
(k_\mu\, dx^\mu)^2$, where the de Sitter metric $d\bar s^2$ is given
by
%%%%%%
\be
d\bar s^2 = -\fft{(1-\lambda\, r^2)\, \Delta_\theta\, dt^2}{
   1+\lambda\, a^2} + \fft{\rho^2\, dr^2}{(1-\lambda\, r^2)(r^2+a^2)}
 + \fft{\rho^2\, d\theta^2}{\Delta_\theta}+
\fft{(r^2+a^2)\, \sin^2\theta\, d\phi^2}{1+\lambda\, a^2}\,,
\ee
%%%%%
with 
%%%%%
\be
\rho^2\equiv r^2+a^2\, \cos^2\theta\,,\qquad
\Delta_\theta\equiv 1+ \lambda\, a^2\, \cos^2\theta\,.  
\ee
%%%%%
The null vector $k_\mu$ is given by
%%%%%
\be
k_\mu\, dx^\mu = \fft{\Delta_\theta\, dt}{1+\lambda\, a^2}
   + \fft{\rho^2\, dr}{(1-\lambda\, r^2)(r^2+a^2)} - 
   \fft{a\, \sin^2\theta\, d\phi}{1+\lambda\, a^2}\,,
\ee
%%%%%
and the function $U$ becomes $U=\rho^2/r$ in this case.

   In $D=5$ we again write the $\mu_i$ coordinates in
(\ref{kdsodd})--(\ref{oddcase3}) as $\mu_1=\sin\theta$,
$\mu_2=\cos\theta$, and define $\phi_1=\phi$, $\phi_2=\psi$, $a_1=a$
and $a_2=b$.  The Kerr-de Sitter metric is then written as $ds^2 =
d\bar s^2 +\ft{2M}{U}\, (k_\mu\, dx^\mu)^2$, where the de Sitter
metric $d\bar s^2$ is given by
%%%%%%
\bea
d\bar s^2 &=& - \fft{(1-\lambda\, r^2)\, \Delta_\theta\, dt^2}{
   (1+\lambda\, a^2)(1+\lambda\, b^2)}
 + \fft{r^2\, \rho^2\, dr^2}{(1-\lambda\, r^2)(r^2+a^2)(r^2+b^2)}
+ \fft{\rho^2\, d\theta^2}{\Delta_\theta}\nn\\
&& + \fft{r^2+a^2}{1+\lambda\, a^2}\, \sin^2\theta\, d\phi^2 +
\fft{r^2+b^2}{1+\lambda\, b^2}\, \cos^2\theta\, d\psi^2\,,
\eea
%%%%%
where we define
%%%%%
\be
\rho^2 \equiv r^2 + a^2\, \cos^2\theta + b^2\, \sin^2\theta\,,\qquad
\Delta_\theta \equiv 1 +\lambda\, a^2\, \cos^2\theta + 
          \lambda \, b^2\, \sin^2\theta\,.
\ee
%%%%%
The null vector $k_\mu$ is given by
%%%%%
\be
k_\mu\, dx^\mu = \fft{\Delta_\theta\, dt}{(1+\lambda\, a^2)
    (1+\lambda\, b^2)}\,
 + \fft{r^2\, \rho^2\, dr}{ (1-\lambda\, r^2)(r^2+a^2)(r^2+b^2)} 
-\fft{a\, \sin^2\theta\, d\phi}{1+\lambda\, a^2} -
 \fft{a\, \cos^2\theta\, d\psi}{1+\lambda\, b^2}\,.
\ee
%%%%%
The function $U$ is just given by $U=\rho^2$ in this case.

\section{Kerr-de Sitter metrics with $a_i=a$ in $D=2n+1$}

    Here, we consider the case where all the rotation parameters
$a_i$ are set equal in the odd-dimensional Kerr-de Sitter metrics.  From
(\ref{oddcase}), this leads to the following special Kerr-de Sitter
metric $ds^2 = d\bar s^2 + \fft{2M}{U}\, (k_\mu\, dx^\mu)^2$, with
%%%%%
\bea
d\bar s^2 &=& -\fft{(1-\lambda\, r^2)\, dt^2}{1+\lambda\, a^2} 
  + \fft{r^2\, dr^2}{(1-\lambda\, r^2)(r^2+a^2)} 
 + \fft{r^2 + a^2}{1+\lambda\, a^2}\, 
   \sum_{i=1}^n(d\mu_i^2 + \mu_i^2 \, d\phi_i^2)\,,\label{oddequal}\\
k_\mu \, dx^\mu &=& \fft{1}{1+\lambda\, a^2}\, \Big(dt - a\sum_{i=1}^n
   \mu_i^2\, d\phi_i \Big)+ \fft{r^2\, dr}{(1-\lambda\, r^2)(r^2+a^2)}\,,
\label{oddequal2}\\
U &=& (r^2 + a^2)^{n-1}\,.
\eea
%%%%%

   The solution can be recast in an even simpler form by noting that
the terms 
%%%%%
\be
 \sum_{i=1}^n(d\mu_i^2 + \mu_i^2 \, d\phi_i^2)
\ee
%%%%%
in the metric (\ref{oddequal}) describe a unit $(2n-1)$-sphere, which
can be viewed as the Hopf fibration over $\CP^{n-1}$, with the metric
written as
%%%%%
\be
(d\psi + A)^2 + d\Sigma_{n-1}^2\,,
\ee
%%%%%
where $\psi$ has period $2\pi$, $d\Sigma_{n-1}^2$ is the canonically
normalised Fubini-Study metric on $\CP^{n-1}$ (\ie with $R_{ij} = 2n\,
g_{ij}$), and $A$ is a local potential for the K\"ahler form $J=\ft12
dA$ on $\CP^{n-1}$.  It is easily seen that $\sum_i \mu_i^2\, d\phi_i$
is equal to $(d\psi + A)$, and so we may re-express the de Sitter
metric $d\bar s^2$ (\ref{oddequal}) and the null vector $k_\mu$
(\ref{oddequal2}) as
%%%%%
\bea
d\bar s^2 \!\!\! &=& \!\!\! -\fft{(1-\lambda\, r^2)
   \, dt^2}{1+\lambda\, a^2} 
 + \fft{r^2\, dr^2}{(1-\lambda\, r^2)(r^2+a^2)} \,
 + \fft{r^2+a^2}{1+\lambda\, a^2}\,
\Big[ (d\psi+A)^2 + d\Sigma_{n-1}^2\Big] \,,\label{fsodd}\\
k_\mu\, dx^\mu \!\!\! &=& \!\!\!
  \fft{1}{1+\lambda\, a^2}\, [dt -a\, (d\psi+A)] 
      + \fft{r^2\, dr}{(1-\lambda\, r^2)(r^2+a^2)}\,.
\eea
%%%%%

   It is worth remarking that the Fubini-Study metric in (\ref{fsodd})
can be replaced by any Einstein-K\"ahler metric with the same scalar
curvature, and with K\"ahler form $J=\ft12 dA$, giving us an infinity of
further possibilities for generalised Kerr-de Sitter metrics $ds^2 =
d\bar s^2 + 2M\, (r^2+a^2)^{1-n} \, (k_\mu\, dx^\mu)^2$.

\section{Kerr-de Sitter metrics with $a_i=a$ in $D=2n$}

    A similar simplification arises if we set the $(n-1)$ rotation
parameters equal in the even-dimensional Kerr-de Sitter metrics.
Writing $\mu_i=(\mu_\a,\mu_n)$, with $1\le\a\le n-1$, and 
%%%%%
\be
\mu_\a = \nu_\a\, \sin\theta\,,\qquad \mu_n=\cos\theta\,,
\qquad  \hbox{where}\qquad \sum_{\a=1}^{n-1} \nu_\a^2=1\,,
\ee
%%%%%
we find that upon setting
$a_\a=a$ in (\ref{kdseven})--(\ref{oddcase3}) the Kerr-de Sitter
metric $ds^2= \bar ds^2 + \fft{2M}{U}\, (k_\mu\, dx^\mu)^2$ is given by
%%%%%
\bea
d\bar s^2 &=& -\fft{(1-\lambda\, r^2)\Delta_\theta\, dt^2}{1+\lambda\, a^2}
 + \fft{\rho^2\, dr^2}{(1-\lambda\, r^2)(r^2+a^2)} + 
\fft{\rho^2\, d\theta^2}{\Delta_\theta} \nn\\
&& \qquad + \fft{(r^2+a^2)\,\sin^2\theta}{1+\lambda\, a^2}\, 
\sum_{\a=1}^{n-1}
(d\nu_\a^2 + \nu_\a^2\, d\phi_\a^2)\,,\\
k_\mu\, dx^\mu &=& \fft{1}{1+\lambda\, a^2}\Bigl(\Delta_\theta\, dt 
 -  a\, \sin^2\theta \, \sum_{\a=1}^{n-1}\, \nu_\a^2\, d\phi_\a
\Bigr) + \fft{\rho^2\, dr}{(1-\lambda\, r^2)(r^2+a^2)}\,,\\
U &=& \fft{\rho^2}{r}\, (r^2+a^2)^{n-2}\,,
\eea
%%%%%
where we define 
%%%%%
\be
\rho^2\equiv r^2+a^2\, \cos^2\theta\,,\qquad
\Delta_\theta \equiv 1+ \lambda\, a^2\, \cos^2\theta\,.
\ee
%%%%%

   We can analogously re-express this in terms of a complex projective
space (which in this case is $\CP^{n-2}$), giving
%%%%%%
\bea
d\bar s^2 &=& -\fft{(1-\lambda\, r^2)\Delta_\theta\, dt^2}{1+\lambda\, a^2}
 + \fft{\rho^2\, dr^2}{(1-\lambda\, r^2)(r^2+a^2)} + 
\fft{\rho^2\, d\theta^2}{\Delta_\theta} \nn\\
&& \qquad + \fft{(r^2+a^2)\, \sin^2\theta}{1+\lambda\, a^2}\, 
\Bigl[ (d\psi + A)^2
   +d\Sigma_{n-2}^2\Bigr]\,,\label{fseven}\\
k_\mu\, dx^\mu &=& \fft{1}{1+\lambda\, a^2}\, [\Delta_\theta\, dt
 - a\, \sin^2\theta\, (d\psi + A)] + 
\fft{\rho^2\, dr}{(1-\lambda\, r^2)(r^2+a^2)}\,.
\eea
%%%%%

   Again, the Fubini-Study metric in (\ref{fseven})
can be replaced by any Einstein-K\"ahler metric with the same scalar
curvature, and with K\"ahler form $J=\ft12 dA$, giving us an infinity of
further generalised Kerr-de Sitter metrics $ds^2 =
d\bar s^2 + 2M\,r \,  (r^2+a^2)^{2-n}\, \rho^{-2} \, (k_\mu\, dx^\mu)^2$.

\section{Kerr-de Sitter metrics with a single rotation parameter}

   Consider first the case when the spacetime dimension is odd, $D=2n+1$.
When there is a single non-vanishing rotation parameter, we may, without
loss of generality, choose it to be $a_1$.  Splitting the $i$ index as
$i=(1,\alpha)$, we define
%%%%%
\bea
\mu_1 &=& \sin\theta\,, \qquad \mu_\a = \nu_\a\, \cos\theta\,,\qquad
   \hbox{where}\qquad \sum_{\a=2}^n \nu_\a^2=1\,,\nn\\
a_1 &=& a\,,\qquad a_\a=0\,,\qquad \phi_1=\phi\,.
\eea
%%%%%
It is then straightforward to see that the $D=2n+1$ Kerr-de Sitter
metrics, given by (\ref{kdsodd})--(\ref{oddcase3}), reduce to $ds^2 =
d\bar s^2 + \fft{2M}{U}\, (k_\mu\, dx^\mu)^2$ with
%%%%%
\bea
d\bar s^2 &=& -\fft{(1-\lambda\, r^2)\, \Delta_\theta\, dt^2}{
   1+\lambda\, a^2} + \fft{\rho^2\, dr^2}{(1-\lambda\, r^2)(r^2+a^2)} 
 + \fft{\rho^2}{\Delta_\theta}\, d\theta^2\nn\\
&& \qquad +
\fft{(r^2+a^2)\, \sin^2\theta\, d\phi^2 }{1+\lambda\, a^2} +
    r^2\, \cos^2\theta\, d\Omega^2_{2n-3}\,,\\
k_\mu\, dx^\mu &=& \fft{\Delta_\theta\, dt}{1+\lambda\, a^2}
   - \fft{\rho^2\, dr}{(1-\lambda\, r^2)(r^2+a^2)} - 
   \fft{a\, \sin^2\theta\, d\phi}{1+\lambda\, a^2}\,,\label{onek}\\
U&=& \rho^2\, r^{2n-2}\,,
\eea
%%%%%
where $d\Omega^2_{2n-3}= \sum_{\a} (d\nu_\a^2 + \nu_\a^2\, d\phi_\a^2)$ is
the unit metric on $S^{2n-3}$, and
%%%%%
\be
\rho^2\equiv r^2+a^2\, \cos^2\theta\,,\qquad 
\Delta_\theta\equiv 1+ \lambda\, a^2\, \cos^2\theta\,.
\ee
%%%%%
The metric $ds^2$ is equivalent to the single-rotation Kerr-de Sitter 
metric given in \cite{hawhuntay}, rewritten in Kerr-Schild coordinates.

   The discussion for the even-dimensional case, $D=2n$, is almost 
identical.  Specialising (\ref{kdseven})--(\ref{evencase3}) by setting
$a_1=a$, $a_\a=0$ for $2\le\a\le n-1$, we find that the single-parameter
Kerr-de Sitter metric takes the same form as in the odd-dimensional 
case above, with $k_\mu$ given by (\ref{onek}), except that now we have
\bea
d\bar s^2 &=& -\fft{(1-\lambda\, r^2)\, \Delta_\theta\, dt^2}{
   1+\lambda\, a^2}+ \fft{\rho^2\, dr^2}{(1-\lambda\, r^2)(r^2+a^2)} 
 + \fft{\rho^2\, d\theta^2}{\Delta_\theta}\nn\\
&& \qquad +
\fft{(r^2+a^2)\, \sin^2\theta\, d\phi^2}{1+\lambda\, a^2} +
    r^2\, \cos^2\theta\, d\Omega^2_{2n-4}\,,\\
U &=& \rho^2\, r^{2n-3}\,.
\eea
%%%%%
Again, this is equivalent to the single-rotation Kerr-de Sitter 
metric given in \cite{hawhuntay}, rewritten in Kerr-Schild coordinates.

\section{Unified Alternative Boyer-Lindquist Form of the Metrics}

Here we give a unified form for the Kerr-de Sitter metrics in all
dimensions, even or odd.  Furthermore, we present it in a somewhat
simpler form than the metric has in the ``standard'' Boyer-Lindquist
coordinates of (\ref{blodd}) and (\ref{bleven}).

  We begin by defining the ``evenness'' integer
%%%
\be
\ep \equiv n-N \equiv [D/2]-[(D-1)/2] = (D-1) \ {\rm mod} \ 2 \,,
\ee
%%%
which is 1 for even $D$ and 0 for odd $D$.  The number of latitudinal
coordinates $\mu_i$ is then $n=N+\ep$, where $N$ is the number of
azimuthal angular coordinates $\phi_i$, $\varphi_i$, or (in this
Appendix) $\hat{\varphi}_i$.  Since the latitudinal coordinates obey
the one constraint (\ref{constraint}), $\sum_i^{N+\ep}\mu_i^2 = 1$,
there are actually $N+\ep-1$ independent latitudinal coordinates.  This
is the same as the number $N$ of azimuthal angular coordinates in even
dimensions $D=2N+2=2n$, but one less in odd dimensions $D=2N+1=2n+1$.
In all cases $D=2N+\ep+1$, which is the sum of the single time
coordinate $t$ or $\tau$, the one radial coordinate $r$, the $N+\ep-1$
independent latitudinal coordinates $\mu_i$, and the $N$ azimuthal
angular coordinates $\phi_i$, $\varphi_i$, or $\hat{\varphi}_i$.

   In this Appendix, instead of the second coordinate transformation
of (\ref{coordtrans}), we shall use
%%%%%
\be
d\phi_i = d\hat{\varphi}_i +
\fft{2M\,a_i\, dr}{(r^2 + a_i^2)(V-2M)}\,,\label{coordtrans2}
\ee
%%%%%
so $\hat{\varphi}_i = \varphi_i - \lambda a_i d\tau$ in terms of the
$\varphi_i$ used in the alternative Boyer-Lindquist coordinates of
Section 3.  Then the Kerr-de Sitter metrics become, in both odd and
even dimensions,
%%%
\bea
ds^2 &=& - W\, (1 -\lambda\, r^2)\, d\tau^2
 + \fft{2M}{VF}\Bigl(W\,d\tau
 - \sum_{i=1}^N \fft{a_i\, \mu_i^2\, d\hat{\varphi}_i}
  {1 + \lambda\, a_i^2}\Bigr)^2
 + \sum_{i=1}^N \fft{r^2 + a_i^2}{1 + \lambda\, a_i^2}\,\mu_i^2\,
    d\hat{\varphi}_i^2 \nn\\
&&
 + \fft{VF\, dr^2}{V-2M}
 + \sum_{i=1}^{N+\ep} \fft{r^2 + a_i^2}{1 + \lambda\, a_i^2}\, d\mu_i^2
 + \fft{\lambda}{W\, (1-\lambda r^2)}
    \Bigl( \sum_{i=1}^{N+\ep} \fft{r^2 + a_i^2}{1 + \lambda\, a_i^2}
    \, \mu_i\, d\mu_i\Bigr)^2 \,,\label{bl}
\eea
%%%
where
%%%%%
\be
W \equiv \sum_{i=1}^{N+\ep} \fft{\mu_i^2}{1+\lambda\, a_i^2}\,,\qquad
\ee
\be
V\equiv r^{\ep-2}\, (1-\lambda\, r^2)\, \prod_{i=1}^N (r^2 + a_i^2)\,,\qquad
F\equiv \fft{1}{1-\lambda\, r^2}\, \, 
  \sum_{i=1}^{N+\ep} \fft{r^2 \, \mu_i^2}{r^2+a_i^2}\,.\label{WVFdef}
\ee
%%%%%

In this alternative Boyer-Lindquist form of the Kerr-de Sitter metric,
the null 1-form is
%%%%%
\be
k_\mu\, dx^\mu = \fft{VF\, dr}{V-2M} \, + \,
W\,d\tau\,-\,\sum_{i=1}^N \fft{a_i\,\mu_i^2\,
d\hat{\varphi_i}}{1+\lambda\,a_i^2}
\,,
\ee
%%%%%
and the corresponding null vector is
%%%%%
\be
k^\mu\, \del_\mu = \fft{\del}{\del r} -\fft{V}{V-2M}\,\left(
 \fft{1}{1-\lambda\, r^2}\, \,\fft{\del}{\del \tau} \,
 +\,\sum_{i=1}^N\,\fft{a_i}{r^2+a_i^2}\fft{\del}{\del \hat{\varphi}_i} \right) 
  \,.
\ee
%%%%%

\end{document}